\documentclass[acmsmall]{acmart}

\AtBeginDocument{%
  }

\setcopyright{acmcopyright}
\copyrightyear{2023}
\acmYear{2023}
\acmDOI{10.1145/1122445.1122456}

\acmJournal{JACM}
\acmVolume{37}
\acmNumber{4}
\acmArticle{111}
\acmMonth{8}

\usepackage{multirow}
\usepackage{pifont}
\graphicspath{ {./images/} }
\usepackage[flushleft]{threeparttable} 
\usepackage{makecell} 
\usepackage{float}
\usepackage{longtable} 
\usepackage{setspace}
\usepackage{soul}

\begin{document}

\title{Addressing Interpersonal Harm in Online Gaming Communities: The Opportunities and Challenges for a Restorative Justice Approach}

\renewcommand{\shorttitle}{Addressing Interpersonal Harm in Online Gaming Communities}

\author{Sijia Xiao}
\affiliation{%
  \institution{University of California, Berkeley}
  \city{Berkeley}
  \country{USA}}
\email{xiaosijia@berkeley.edu}

\author{Shagun Jhaver}
\affiliation{%
  \institution{Rutgers University}
  \city{New Jersey}
  \country{USA}}
\email{shagun.jhaver@rutgers.edu}

\author{Niloufar Salehi}
\affiliation{%
  \institution{University of California, Berkeley}
   \city{Berkeley}
  \country{USA}}
\email{nsalehi@berkeley.edu}


\begin{abstract}
Most social media platforms implement content moderation to address interpersonal harms such as harassment. Content moderation relies on offender-centered, punitive approaches, e.g., bans and content removal. We consider an alternative justice framework, restorative justice, which aids victims in healing, supports offenders in repairing the harm, and engages community members in addressing the harm collectively. To assess the utility of restorative justice in addressing online harm, we interviewed 23 users from Overwatch gaming communities, including moderators, victims, and offenders; such communities are particularly susceptible to harm, with nearly three quarters of all online game players suffering from some form of online abuse. 
We study how the communities currently handle harm cases through the lens of restorative justice and examine their attitudes toward implementing restorative justice processes. Our analysis reveals that cultural, technical, and resource-related obstacles hinder implementation of restorative justice within the existing punitive framework despite online community needs and existing structures to support it. We discuss how current content moderation systems can embed restorative justice goals and practices and overcome these challenges. 

\end{abstract}



\received{2 November 2022}
\received[revised]{9 January 2023}
\received[accepted]{11 March 2023}

\maketitle


\section{introduction} \label{sec:intro}

Social media platforms frequently address online interpersonal harm, such as harassment, through content moderation; this  involves reviewing user-submitted content for appropriateness  and sanctioning contributors that violate the platform's rules \cite{gillespie_custodians_2018, roberts_behind_2019}. However, despite efforts in research and industry to improve moderation practices in recent years, the number of people experiencing severe forms of harassment continues to grow.
In 2014, 15\% of Americans reported experiencing severe harassment, including physical threats, stalking, sexual harassment, and sustained harassment \cite{duggan2014online}. That number grew to 18\% in 2017 and 25\% in 2021. Further, many people report simultaneously experiencing multiple forms of severe harassment \cite{duggan_online_2017, vogels2021state}. Research shows that online harms are insufficiently addressed by platforms' and communities' current approaches \cite{duggan2013,citron_addressing_2015,lenhart2016online,warzel2016honeypot}, and 32\% of Americans say that social media companies are doing a poor job at addressing online harassment on their platforms \cite{vogels2021state}. Alternative approaches are desperately needed, but what principles should guide them, and how would they work in practice?


\textit{Restorative justice} is a framework that argues for repairing harm and restoring individuals and communities after harm has occurred. In this paper, our goal is to \textit{draw from restorative justice philosophy and practice to study how an online gaming community currently addresses---and might alternatively address---interpersonal harm}. We focus on restorative justice here because it has an established offline practice and has been successfully institutionalized to address harm in other contexts, such as schools and prisons
\cite{kuo2010empirical, latimer_effectiveness_2005, bazemore2001comparison}. In recent years, the HCI and CSCW communities have also explored its utility in addressing online harm \cite{blackwell2018online, schoenebeck2020drawing, kou2021punishment}, and we build upon these efforts, as well. 

Restorative justice focuses on providing care, support, and in other ways meeting people's needs after harm has occurred. It has three major principles: (1) identify and address the victim’s needs related to the harm, (2) support the offender in taking accountability and working to repair the harm, and (3) engage the community in the process to support victims and offenders and heal collectively \cite{mccold2000toward, zehr2015little}. In practice, restorative justice addresses harm differently than more common punitive models. The main tool for action in a punitive justice model, as embodied in content moderation, is \textit{punishing} the rule violator. 
In contrast, in restorative justice it is \textit{communication} among the harmed person, the offender, and the community. For instance, in a common restorative justice practice called a \textit{victim-offender conference}, the victim and offender meet to discuss the harm and how to address it under the guidance of a facilitator; interested community members are also invited to join this conversation since the conference aims to address the needs and obligations of all three parties involved. 
A follow-up process may include apologies or community service by the offenders \cite{zehr2015little}.

This paper uses the three restorative justice principles described above and its common practices (e.g., the victim-offender conference) as a vehicle to study the perspectives and practices of victims, offenders, and moderators during instances of online harm. First, we use the principles to 
evaluate current practices for addressing interpersonal harm and identify the potential need for restorative justice practices. We focus on the experiences of victims, offenders, and moderators, who are key stakeholders and participants in restorative justice conferences (with the moderator acting as facilitator).
Second, we 
use the victim-offender conference to identify the benefits and challenges of practicably implementing restorative justice practices in an online setting.

We study harm cases in the Overwatch gaming community, which spans two major platforms : the Overwatch platform on which the game is played \footnote{https://playoverwatch.com} and the Discord platform  \footnote{https://discord.com} on which gaming discussions, teammate selection, and match organization occur. Online gaming communities have long suffered from severe and frequent incidents of online harm \cite{adinolf2018toxic,hilvert2020m, beres2021don}. 
Our analysis of Overwatch, a multi-player game, lets us explore such harm in the context of different types of  user relationships, including competition and collaboration. 
We interviewed self-identified victims (people who have been harmed), offenders (people who have harmed others), and moderators who dealt with the cases being discussed. Our interview protocol resembles the restorative justice practice of \textit{pre-conferencing}, which is used to learn people's history and preferences, explain restorative justice to them, and evaluate the appropriateness of holding a victim-offender conference \cite{zehr2015little}. Additionally, given that restorative justice has been chiefly developed through practice\cite{van2014restoring}, we deepened our understanding of its principles and practices by conducting two interviews with 
offline restorative justice practitioners.

We find that current, punitive online moderation processes do not effectively stop the perpetuation of harm. First, content moderation is offender-centered and does not address victims' needs, such as receiving support or healing from harm. Though victims may report individual offenders, they continue to receive harm in a community where abuse is prevalent. Second, content moderation directs offenders' attention to the punishment they receive instead of the damage they cause. When punishment is ineffective, as is often the case, there are no alternative ways to hold offenders accountable. Finally, community members with a punitive mindset may further perpetuate harm by not acknowledging the harmed person's experiences or reacting punitively toward perpetrators or victims, particularly when harm cases are complex and layered. 

Our findings show that some current moderation practices align with restorative justice, and a few participants have attempted to implement restorative justice practices in their own online communities. Some victims and offenders also expressed needs that align with restorative justice values. However, applying restorative justice online is not straightforward: there are structural, cultural, and resource-related obstacles to implementing a new approach within the existing punitive framework. 
We elaborate on the potential challenges of implementing restorative justice online and propose ways to design and embed its practices in online communities. 

Our work contributes to a growing line of research that applies alternative justice frameworks to address online harm \cite{dencik2017towards, citron_addressing_2015, schoenebeck2020drawing,musgrave2022experiences, hasinoff2020promise, hasinoff2022scalability, salehi2020donoharm}. By evaluating the moderation practice of the Overwatch gaming community through a restorative justice lens and identifying key stakeholders' preferences for the justice framework, our work sheds light on ways to address online harm that go beyond simply maintaining healthy content 
and working within a perpetrator-oriented model. We highlight how restorative justice has the potential to reduce the continuance of harm and improve community culture in the long run.

\section{Related Work}
We now review related work on online harm and content moderation. We also describe the restorative justice framework and its current applications.

\subsection{Online Harm and Content Moderation}
Online harm can refer to a myriad of toxic behaviors, such as public shaming \cite{ronson2015one}, trolling \cite{cheng2015antisocial}, and invasion of privacy \cite{lenhart2016online}. Our research focuses on interpersonal harm (as opposed to self-directed or collective harm), which is the harm that occurs in the interaction between two or several individuals \cite{krug2002world}. Most of the interpersonal harm we study falls under Duggan's description of online harassment, which includes six categories: offensive name-calling, purposeful embarrassment, physical threats, sustained harassment, stalking, and sexual harassment \cite{duggan_online_2017}.

Online gaming communities are particularly susceptible to harm: nearly three-quarters of all online game players have suffered some form of online abuse \cite{league2019free}. Within the context of online games, ``toxicity'' describes abusive communication (e.g., verbal or textual harassment) as well as disruptive gaming behaviors (e.g., exploiting glitches, sabotaging teammates) \cite{reid2022bad, adinolf2018toxic}. Toxic behaviors hurt the gaming experience of involved players; prior research has found that toxicity during game play can negatively affect individual and team performance \cite{monge2022effects}. 
Even so, it is pervasive and normalized as part of the competitive game culture \cite{adinolf2018toxic,hilvert2020m, beres2021don}.
Notably, this toxicity intersects with harm toward certain groups, for example, females \cite{shaer2017understanding, fox2017women, ruvalcaba2018women, ruotsalainen2018there}, gender minorities \cite{ruberg2018queerness}, and people of color \cite{gray2012intersecting}.
Reid et al. have argued that platforms should offer in-game support tools that immediately address toxicity or help players cope with harm after it occurs 
\cite{reid2022feeling}. We respond to this call by examining how introducing restorative justice tools and processes may address harm in online gaming communities.



Online social media and gaming platforms currently address harm through content moderation, which usually involves punitive measures such as removing content, muting, or banning offenders \cite{gillespie_custodians_2018, roberts_behind_2019}. Online platforms' moderators can be volunteers who are  platform users or commercial content moderators hired by social media companies \cite{seering_moderator_2019,roberts_behind_2019}. In recent years, social media platforms have also begun using automated, AI-based tools such as bots to help enact moderation \cite{binns2017like,jhaver2023personalizing,kiene2020uses}. 

As the most widely applied model of addressing online harm, content moderation faces many implementation challenges and is criticized for its limits in building healthy communities. One key challenge is the sheer amount of labor required: as social media platforms grow, moderators must address increasing numbers of harm cases. Another challenge is the emotional labor required to moderate potentially upsetting antisocial content \cite{roberts_behind_2019, dosono2019moderation, wohn2019volunteer}. At the same time, moderation is not always efficient and effective in addressing harm: content policies and their implementations have often failed to sufficiently remove disturbing material like fake news (a colloquial term for false or misleading content presented as news) \cite{allcott2017}, alt-right trolls \cite{nagle2017kill,romano2017} and revenge porn \cite{citronfranks2014,vanian2017}. 

Researchers have examined ways to improve the current moderation processes, for example, through setting positive examples and social norms \cite{seering2017shaping,chandrasekharan2018norms}, providing explanations for post removals \cite{jhaver2019survey,jhaver2019transp}, 
supporting creators in configuring automated moderation filters \cite{jhaver2022wordfilters}, 
and placing warning labels on inappropriate content \cite{chandrasekharan2022quar,morrow2022emerging}.
We add to these efforts by taking a user-centered approach that interrogates how victims, offenders, and community members experience the current content moderation approach in addressing harm through the perspective of restorative justice and identifying opportunities for improvements.

Prior research has found that victims often have needs that content moderation does not address, e.g., the need to make sense of the harm \cite{to2020they, xiao2022sensemaking}, receive emotional support and validation \cite{thomas2022s}, or transform the online environment to prevent harm from happening \cite{xiao2022sensemaking}. Researchers have also developed tools to support victims beyond content moderation, such as tools for social support \cite{mahar_squadbox:_2018, blackwell2017classification} or collecting and documenting evidence \cite{sultana2021unmochon, goyal2022you}. These tools are often external to the platform on which the harm occurs. 
We examine how in-house, community-centered restorative processes, which allow a deeper reflection on online harms, may help address victims' needs.

In recent years, researchers have begun asking questions about the role that social media platforms play in the realization of values important to public discourse, such as freedom of expression, transparency, protection from discrimination, and personal security and dignity \cite{suzor2018evaluating,gorwa2019platform,helberger2018governing,suzor2018digital,suzor2019we,myers_west_censored_2018}. Many scholars have offered ethical frameworks to guide platforms' content moderation efforts. Perhaps the most prominent framework for ethical conduct in social media moderation is the Santa Clara Principles \cite{santaclaraprinciples}, which outline ``minimum levels of transparency and accountability'' and propose three principles for providing meaningful due process: (1) publishing data about removed posts, (2) notifying users affected by moderation, and (3) granting moderated users a right to appeal. Shannon Bowen applies Kantian deontology to social media content management decisions, advocating that public relations practitioners rely on universal principles such as dignity, fairness, honesty, transparency, and respect when communicating via social media \cite{bowen2013using}.
%
%
A variety of other frameworks, originally developed in offline settings, usefully inform the ethics of content moderation \cite{carlson2020you}. Yet, restorative justice is an especially informative framework for our narrow focus on addressing online harm because it prioritizes individuals and relationships rather than content; it therefore forms the focal point of our research.

We contribute to an emerging line of research that examines the values of a diverse range of stakeholders of online platforms to inform governance practices. For example, Schoenebeck et al. inquired about various victims' preferences for moderation measures based on frameworks such as restorative justice, economic justice, and racial justice \cite{schoenebeck2020drawing}. 
Marwick developed an explanatory model of networked harassment by interviewing people who had experienced harassment and Trust \& Safety workers at social platforms \cite{marwick2021morally}.
Jhaver et al. conceptualized the distinctions between controversial speech and online harassment by talking to both victims and alleged perpetrators of harassment \cite{jhaver2018view,jhaver2018blocklists}.
Supporting such inquiries, Helberger et al. argues that the realization of public values should be a collective effort of platforms, users, and public institutions \cite{helberger2018governing}. 
Our research builds on this line of work by considering victims, offenders, and community members’ needs and values in harm reparation.

\subsection{Restorative Justice}
Punitive justice is often most familiar to us since it is the culturally dominant way we deal with harm. Therefore, to explain restorative justice, we first contrast it with punitive justice. Next, we detail common practices in restorative justice and their application in offline scenarios.

In Western cultures, the dominant model for justice when harm occurs involves punishing the offender \cite{szablowinski2008punitive}. This punitive justice model holds that harm violates rules and offenders should suffer in proportion to their offense \cite{garland2012punishment}. The central focus of this model is on punishing and excluding the offender. However, the victim's concerns about the effects of the offense are rarely taken into account. Further, this model does not help offenders become aware of the negative impact they cause and take accountability to repair the harm \cite{szablowinski2008punitive}. 

Analyzing how harm is addressed in early MUD (multi-user dungeon) communities, Elizabeth Reid observed that ``Punishment on MUDs often shows a return to the medieval. While penal systems in the western nations...have ceased to concentrate upon the body of the condemned as the site for punishment, and have instead turned to `humane' incarceration and social rehabilitation, the exercise of authority on MUDs has revived the old practices of public shaming and torture'' \cite{reid1999hierarchy}. Two decades later, the public spectacle of punishment still prevails on current digital platforms.

\subsubsection{Restorative Justice Principles}
Restorative justice provides an alternative way to address harm. It supports the belief that harm is a violation of people and relationships rather than merely a breach of rules \cite{van2016overview}. It puts victims at the center of the process and seeks to repair the harm they suffer from the offense. Restorative justice has three major principles \cite{mccold2000toward}: (1) provide support and healing for victims, (2) help offenders realize the consequences of their wrong-doing and repair the harm, and (3) encourage communities to provide support for both victims and offenders and to heal collectively. Multiple levels of communities can be primarily affected by harm, including the local community where the harm occurs or the broader society \cite{mccold2000toward}. Our study addresses victims and offenders in the Overwatch gaming community.


\subsubsection{Restorative Justice Practices}
Restorative justice has been successfully applied in a myriad of settings, such as the criminal justice system, schools, and workplaces \cite{van2016overview,wood2016four}. When formalized within an organization, restorative justice processes can be embedded within a punitive justice system to use on selected types of harm cases \cite{van2016overview,wood2016four}. Those who do not want to proceed using this approach or cannot reach a consensus during the restorative justice process are directed to the punitive justice system \cite{bazemore2001comparison} for redress.

A widely used practice in restorative justice is the \textit{victim-offender conference} \cite{zehr2015little}. Here, victims and offenders meet with a restorative justice facilitator to discuss three core questions: (1) what has happened? (2) who has been affected and how? and (3) what is needed to repair the harm? The facilitator mediates this process to ensure that victims and offenders have equal footing and helps move the parties toward reaching a consensus. Other forms of restorative justice meetings, such as family-group conferences, include additional community members, such as family and friends of victims and offenders \cite{zehr2015little, bazemore2001comparison}. These practices embed values and principles of restorative justice to meet the needs of all parties involved, including the victim, offender, and community members. In this paper, we use victim-offender conferences as an exemplar to inquire about participants' preferences for online restorative justice practices.

Restorative justice practitioners emphasize the importance of preparation in advance of the victim-offender conference. A restorative justice facilitator first meets separately with the offender and the victim before the conference in a process called a \textit{pre-conference} \cite{zehr2015little}. During these meetings, the facilitator introduces the restorative justice framework to the victim and offender and asks them the same questions they will be asked during the conference. After these meetings, the victim-offender conference happens only when both parties agree to meet voluntarily to repair the harm. Additionally, the facilitator acts as the gatekeeper to determine whether victims and offenders can meet to reach a desired outcome without causing more harm. In both pre- and victim-offender conferencing, the facilitator does not make decisions for victims or offenders about how to address the harm but guides them to reflect on the harm through the restorative justice framework \cite{bolitho2017science}. For this research, the first author received training and acted as a facilitator in pre-conferencing with the participants to ask about their attitudes toward engaging in a restorative justice process to address online harm.

\subsubsection{Restorative Justice Outcomes}
Restorative justice does not encourage punishment as the desired outcome. Instead, it focuses on the obligation to repair harm and heal those hurt \cite{zehr2015little}. Possible outcomes of a restorative justice conference include an apology from the offender or an action plan that the offender will carry forward, e.g., doing community service or attending an anger management course \cite{daly_restorative_2002}. This framework acknowledges that it can be difficult, or in some cases even impossible, to fully restore the situation or repair the damage. However, symbolic steps, including acknowledgment of impact and an apology, can help victims heal and offenders learn and take accountability \cite{pranis2015little}. 
When restorative justice is embedded within a punitive system, its outcome can inform the punishment decision in some cases \cite{van2016overview}. For example, when victims and offenders can reach a consensual outcome in a restorative justice process, offenders may receive a reduction or exemption from a punitive process. In other cases, a restorative justice process may run in parallel and have little influence on the punitive process \cite{van2016overview}.

Here, an example may be instructive. A high school in Minnesota was dealing with problems of drug and alcohol abuse. The school held a conference that gathered the offender (a student who had used drugs on school grounds), victims (affected students), and community members (faculty and staff). The offender first shared her story and the reasons for her actions. She also took the opportunity to ask for forgiveness. Other members of the conference then expressed how they were affected by the offender's behavior and jointly discussed solutions. The outcome was that the offender became aware of the effects of her actions and agreed to go through periodic checks to monitor her continued sobriety \cite{karp_restorative_2001}.

Though practicing restorative justice benefited the affected parties in the case above, does it always succeed?
Latimer et al. empirically analyzed existing literature on the effectiveness of restorative justice. 
They found that in-person restorative justice programs successfully reduced offender recidivism and increased victims' satisfaction with the process and the outcome \cite{latimer_effectiveness_2005}. 
However, these positive findings were tempered by the self-selection bias inherent in restorative justice practices. Since it is a voluntary process, those who choose it may benefit more than others  \cite{latimer_effectiveness_2005}. Restorative justice also requires commitment at the administrative level \cite{gavrielides2017restorative} and time and labor for the parties involved \cite{zehr2015little}. In this research, we examine the potential of applying restorative justice to online settings for addressing harm. The unique characteristics of online communities---such as anonymity \cite{lapidot2012effects}, lack of social cues \cite{donath_identity_2002}, and weak social ties \cite{haythornthwaite2002strong}---create new challenges and opportunities for adapting restorative justice practices online. 

Implementing  restorative justice values to address online harm is gaining traction in the research community. Blackwell et al. first introduced restorative justice in the content moderation context \cite{blackwell2018online}. Schoenebeck et al. conducted a large-scale survey study, showing that restorative approaches, such as using apologies to mitigate harm, were strongly supported by participants \cite{schoenebeck2020drawing}. West proposed that education may be more effective than punishment for content moderation at scale \cite{myers2018censored}. 

In the context of online gaming communities, Kou argued that permanent bans produce stereotypes of the most toxic community members and are ineffective over the long term; he recommended that online communities use a restorative justice lens instead of dispensing bans to re-contextualize toxicity and reform members into becoming well-behaved contributors \cite{kou2021punishment}. Hasinoff and Schneider examined the tension between online platforms' emphasis on scalability and the principles of restorative and transformative justice, which prioritize tailored and individualized harm solutions. They argued that `subsidiarity,' the principle that local social units should have meaningful autonomy within larger systems, may address this tension \cite{hasinoff2022scalability}. We build on this rich line of work by using the Overwatch gaming community as a case study to explore the benefits and challenges of restorative justice for different stakeholders.


\section{Background}
To provide context for our methods and results, we briefly review the two platforms we study, Overwatch and Discord.

\subsection{Overwatch and Its Moderation Practices}
Overwatch is a real-time, team-based video game developed and published by Blizzard Entertainment\footnote{https://www.blizzard.com}. It assigns players to two opposing teams of six. Gamers play in the first-person shooter view and can select from over 30 \textit{heroes} with unique skills. They pair up with random players if they enter the game alone, but they can also choose to pair with selected teammates. During each game, players communicate through the built-in voice chat and text chat functions, but some players also use Discord as an alternative. 
All players are expected to comply with a set of rules laid out in Blizzard's code of conduct \cite{blizzard_blizzards_2020}. For example, these rules instruct, ``You may not use language that could be offensive or vulgar to others,'' and ``We expect our players to treat each other with respect and promote an enjoyable environment.''

Blizzard hires commercial content moderators, who are paid company employees, to regulate its games \cite{roberts_behind_2019}. Though the company shows users a small set of moderation rules in its code of conduct, it is likely that the company has an internal set of more detailed moderation guidelines to help moderators make their decisions \cite{roberts_behind_2019}. 
While moderators do not monitor live games, they handle reports from victims by reviewing game replays and take moderation actions if they determine that users have violated platform rules. Typically, offenders receive a voice chat ban or a temporary or permanent account ban. Offenders receive the decision notification, usually without a detailed explanation. Victims who report the incident usually receive a notice that an action has been taken, but they are not told what the action is. 

\subsection{Discord and Its Moderation Practices}
Discord is a popular instant messaging platform that is widely used by Overwatch gamers. On Discord, users can create their own communities, called \textit{servers}, that contain both text and voice channels for real-time discussions. At the time of this study, more than 2000 Discord servers were active under the tag ``\#Overwatch.'' Overwatch gamers use these servers to discuss the game, find teammates, and organize Overwatch matches.

Discord moderators are volunteer end-users who regulate their communities and screen posts for inappropriate content \cite{jiang_moderation_2019}. Each community creates its own set of moderation rules. Moderators can sanction users by removing their posts, muting them, or banning them either temporarily or permanently. Since moderators and users both have access to public channels in real-time, moderators can actively monitor harm cases on those channels as they occur. They cannot access private channels, but users can report harmful incidents to moderators through private messages. Some communities use automated moderation tools to detect posts containing inappropriate keywords and issue automatic warnings to posters \cite{jiang_moderation_2019}. 
We contribute to the study of Discord moderation by showing how different stakeholders perceive and engage with cases of online harm.

\subsection{The Overwatch Gaming Community Spans Overwatch and Discord}
Like many online communities, the community we study spans multiple platforms \cite{fiesler2020moving}: Overwatch, the gaming platform, and Discord, a discussion platform. 
Though Overwatch pairs up random players if they enter alone, Overwatch gamers frequently use Discord to discuss the game, stay connected, and communicate with one another. Harm can occur both for  players with pre-existing social connections and those who are strangers to one another. Since volunteer moderators participate in the Discord community and are often gamers, they may also be friends with a victim and/or offender in a harm case. In addition, a harm case may initiate in Overwatch but extend to Discord, or vice versa. Our research investigates harm cases on Overwatch, Discord, or both platforms and includes  players with diverse social relationships.
\section{methods}

In total, we interviewed 23 participants from the Overwatch gaming community for this study (Table \ref{table:1}).  To understand Overwatch gamers' perspectives on the restorative justice process, we interviewed victims (those harmed), offenders (those who harm)\footnote{
We use the terms ``victim'' and ``offender'' for brevity and to clarify  participants' roles in specific harm cases. We recognize and agree with calls for eradicating the use of these labels over the long term. Some restorative justice practitioners believe these labels and their meanings are rooted in the punitive justice system, and transformation to restorative justice requires transforming our language. Using ``offenders'' may imply that people are ``inherently bad'' and deserve the condemnation of society \cite{branham_2019}, while using ``victims'' may feel disempowering to some and deny the agency victims should have in restorative justice \cite{gavrielides2017collapsing}. Less popular alternatives to these terms include ``the person who caused harm'' or ``perpetrator'' and the ``person who has been harmed'' or ``survivor.'' Accordingly, the language of ``victim-offender conference'' has also shifted to ``restorative justice circle.''
However, such alternative terms may also not align with the self-image of harmed participants, as we found during our data collection. For this early-stage research, we retain the original terminology to be consistent  with the language participants used.
}, and Discord moderators who dealt with the cases being discussed. 
Some participants fall into more than one of these three groups. 
We could not include Overwatch moderators in our study because they are commercial content moderators \cite{roberts2018digital} who remain anonymous and constrained by non-disclosure agreements. 

Restorative justice primarily evolved through practice rather than as an academic discipline. To more deeply understand how its principles might be applied online, we conducted two additional expert interviews with facilitators from a restorative justice center at the University of California, Berkeley. Further, the first author attended 30 hours of restorative justice training courses to learn how it is practiced in local communities. These interviews and the training helped ground our research in restorative justice values and practices.


\subsection{Recruitment}
We recruited participants using a combination of convenience sampling and snowball sampling \cite{robinson_sampling_2014,biernacki_snowball_1981}. First, we joined multiple Discord communities focused on Overwatch and reached out to moderators, sending private messages to request an interview. After building rapport with moderators through interviews, we asked their permission to publish recruitment surveys in their communities to find victims and offenders of harm. Some moderators referred us to their fellow moderators for interviews and invited us to other Overwatch Discord communities they were involved in. In total, we recruited participants from five Overwatch Discord `server' communities. In addition, we recruited two facilitators for expert interviews from a training program the first author participated in.
We recruited victims and offenders separately through two surveys. In the survey for victims, we described our recruitment criteria as people who have experienced online harm on Overwatch Discord or in an  Overwatch game.  During the interviews, some victims referred us to their friends who have experienced harm or have been banned in the Overwatch gaming community, and we included them as participants. For the second survey, we did not describe participants as ``offenders'' or ``people who have caused harm'' since prior research suggests that people may not want to associate themselves with those categories, especially when there has been no opportunity to discuss what has happened \cite{jhaver2018view,jhaver2018blocklists}. Therefore, we described the recruitment criteria as people who have been warned or banned on Overwatch Discord or in the game. 


In the recruitment surveys, we asked participants to briefly describe a harm case they had experienced. We selected participants from this survey based on the time order of their replies. Additionally, we conducted preliminary data analysis to categorize the types of harm (e.g., on Discord vs Overwatch; between friends/strangers; within the moderation team/between end-users/between end-users and moderators).  We then prioritized participants who had experienced different types of harm for interviews. Table \ref{table:1} describes the demographic information of our participants.

\begin{table}
\fontsize{8}{10}\selectfont
\vspace{-1mm}
\caption{Participants' demographic information. We recruited participants using surveys for Overwatch or Discord users who (1) have been harmed or (2) have been banned or warned. Additionally, we recruited moderators on Discord. We show here the demographic details of each participant and their self-identified role (victim, offender, moderator, or facilitator; marked by `x') in the harm cases they discussed with us. Note that a single person may have multiple roles in a harm case or across different cases. We also recruited two restorative justice facilitators.}
\vspace{-3mm}
\begin{tabular}{ m{0.3cm} m{0.3cm} m{1.34cm} m{1.2cm} m{2.5cm} m{0.9cm} m{0.7cm} m{1cm} m{1.1cm} m{1.2cm}}
\midrule
\textbf{ } & \textbf{Age} \tnote{1} & \textbf {Gender}& \textbf{Race/
ethnicity} & \textbf{Education} & \textbf{Country} & \textbf{Victim} & \textbf{Offender} & \textbf{Moderator} & \textbf{Facilitator}\\
\midrule
P1	&	25	&	Female	&	Asian	&	Master's degree	&	US	&	x	&		&	x	&		\\
P2	&	20	&	Male	&	White	&	Bachelor's degree	&	UK	&		&		&	x	&		\\
P3	&	24	&	Non-binary	&	White	&	Some college	&	UK	&		&		&	x	&		\\
P4	&	20s	&	Female	&	White	&	Associate degree	&	CA	&		&		&	x	&		\\
P5	&	26	&	Female	&	Mixed	&	Master's degree	&	US	&		&		&		&	x	\\
P6	&	27	&	Male	&	Fula	&	Master's degree	&	US	&		&		&		&	x	\\
P7	&	18	&	 --- 	& --- 	&	Some college	&	UK	&		&		&	x	&		\\
P8	&	19	&	Female	&	White	&	Some college	&	UK	&	x	&		&		&		\\
P9	&	20	&	Male	&	White	&	Some college	&	US	&	x	&		&		&		\\
P10	&	20	&	Male	&	Hispanic	&	Some college	&	N/A	&		&		&	x	&		\\
P11	&	21	&	Female	&	White	&	Associate degree	&	US	&	x	&		&		&		\\
P12	&	18	&	Female	&	Mixed	&	Less than high school	&	US	&	x	&		&		&		\\
P13	&	19	&	Female	&	White	&	High school graduate	&	CA	&	x	&		&		&		\\
P14	&	24	&	Female	&	White	&	Bachelor's degree	&	CA	&		&		&	x	&		\\
P15	&	18	&	 --- 	&	White	&	Less than high school	&	NA	&		&		&	x	&		\\
P16	&	25	&	Transgender	&	White	&	Master's degree	&	US	&	x	&		&	x	&		\\
P17	&	18	&	Male	&	Asian	&	Some college	&	US	&	x	&	x	&		&		\\
P18	&	23	&	Male	&	White	&	Bachelor's degree	&	UK	&	x	&		&	x	&		\\
P19	&	30	&	Male	&	White	&	Master's degree	&	UK	&	x	&	x	&	x	&		\\
P20	&	18	&	Male	&	Asian	&	Some college	&	CA	&	x	&	x	&	x	&		\\
P21	&	18	&	Male	&	White	&	Less than high school	&	US	&	x	&	x	&		&		\\
P22	&	20	&	Male	&	White	&	N/A	&	UK	&	x	&	x	&	x	&		\\
P23	&	18	&	Male	&	White	&	High school graduate	&	Ireland	&	x	&		&		&		\\
P24	&	18	&	Male	&	Berbers	&	Less than high school	&	Algeria	&	x	&	x	&	x	&		\\
P25	&	24	&	Female	&	White	&	Bachelor's degree	&	CA	&	x	&		&	x	&		\\
\midrule
 \end{tabular}
  \label{table:1}
  \begin{tablenotes}\footnotesize
      \item[1] Participants' gender and race/ethnicity are self-identified.
    \end{tablenotes}

  \vspace{-6mm}
\end{table}

\subsection{Interview Procedure}
Through \textbf{interviews with victims and offenders}, we wanted to understand both their current experiences with harm cases and their perspectives on a restorative justice process for those cases. We adapted our interview questions from restorative justice pre-conferences, where facilitators meet one-on-one with victims and offenders to solicit their perceptions on using a restorative justice process to address harm \cite{zehr2015little}. During the first stage of the interviews, we asked participants questions about the harm case they had experienced or caused, including how it was handled, its impact, and the need to address the harm. 

During the second stage, we introduced participants to restorative justice principles and the victim-offender conferences. 
We focused on these conferences because they constitute a widely used practice that embeds the core restorative justice principles. We included frequently posed questions in preparation for victim-offender conferences, such as what information they would like to convey in the conference and their expectations and concerns regarding the process. If time allowed, we asked participants to reflect on more than one harmful incident. 
Further, some victims and offenders were also moderators of the community. We first inquired about the harm with their primary self-identified roles and then asked about their secondary role(s).

Our \textbf{interviews with Discord moderators }were intended to assess how they deal with harm in their communities and their attitudes toward using restorative justice to repair the harm. Additionally, since the facilitator is essential in offline restorative justice practices, we explored the possibility of creating a corresponding role in online scenarios. Since moderators have the closest currently existing role to a facilitator,  we sought to learn their perspectives on assuming this role. During interviews with them, we first asked about the harm cases they had handled in their communities and their decision rationales. We then introduced the idea of a victim-offender conference and asked them to (1) reflect on potentially using it as an alternative approach for addressing harm cases they had handled in their communities and (2) share  their thoughts about serving as restorative justice facilitators on those cases.

It is challenging to elicit people's perspectives on a hypothetical process or a process they lack previous knowledge about. To make restorative justice concepts more concrete, we asked participants to imagine a restorative justice process based on actual harm cases they had experienced or handled. Additionally, we answered their follow-up questions 
and corrected any misconceptions we identified in our discussions. We continued analyzing our interview data as we recruited and interviewed more participants. We ceased recruiting when our analysis reached theoretical saturation \cite{charmaz_constructing_2006}.

We conducted \textbf{two expert interviews} with restorative justice facilitators to elicit their insights about using restorative justice in online settings. We introduced these facilitators to Discord and Overwatch moderation mechanisms and described examples of how harm cases were handled based on our interviews with victims and offenders. Here, we stayed close to our raw data and described the harm cases through the perspectives of our participants. The facilitators evaluated the current moderation practices through the restorative justice lens and envisioned the future of restorative justice on Discord and Overwatch. We did not intend to reach theoretical saturation for this population \cite{charmaz_constructing_2006}; we still incorporated these interviews 
because doing so provided valuable insights on how these cases could be alternatively handled in a restorative justice context \cite{salehi2017communicating}.

We conducted our interviews from February to July 2020. The interviews lasted one to two hours each, and participants received compensation from \$25 to \$50 US dollars based on interview duration. We conducted 21 interviews using Discord's voice call function, and two participants (P1, P13) chose to be interviewed using Discord's text chat. Before each interview, we negotiated interview time, which was based on participants' availability and the number of harm cases they wanted to share during the interview. We also conducted two in-person interviews with the facilitators (P5, P6).  Our study was approved by the Institutional Review Board (IRB) at the University of California, Berkeley. 

\subsection{Data Analysis}
We conducted interpretive data analysis on our interview transcripts \cite{charmaz_constructing_2006}. We began with a round of initial coding \cite{saldana2013coding}, applying short phrases as codes to our data line-by-line to keep the codes close to the data. Examples of first-level codes included ``impact of harm,'' ``creating new account,'' and ``banning.'' Next, we conducted focused coding \cite{saldana2013coding} by identifying frequently occurring themes and forming higher-level descriptions; second-level codes included ``notion of justice'' and ``sense of community.'' Coding was done iteratively, where the first author frequently moved between interview transcripts and codes and discussed emergent themes with other authors. After these initial and focused coding rounds, we applied restorative justice principles and values as a lens to guide our interpretations. Finally, we established connections between our themes to arrive at our findings, which we categorized according to participants' roles (offenders, victims, moderators, and facilitators). We coded the expert interviews using the same code book we used for other interviews since it helped us compare facilitators' views of harm to other participants' opinions during the analysis.

\subsection{Methodological Limitations}
We used convenience and snowball sampling to recruit participants in a 
single gaming community. In addition, our research examines harm from an interpersonal perspective. As a result, the experiences participants report and the solutions we design may not be representative of, or applicable to, all forms of gaming communities or online harm. Further, though Overwatch has a multinational and multicultural user base, most participants were English speakers from Western cultures. 
Finally, our recruitment method did not let us recruit offenders who were neither warned nor banned. 

As researchers, we lack a full picture of what occurred in the harm cases we studied. To this end, we adopted Whitney Phillips's approach \citep{phillips2015we}, which involved observing how our participants presented themselves to us and drawing conclusions from their performance, which may have been choreographed. As a result, we present our findings as subjective perspectives of Overwatch members rather than as objective truths.
During the interviews, all victims believed that they were the party that had received harm; all offenders believed they had caused harm, but some assumed  only partial accountability since they were also harmed in the process. 
We analyzed our interviews based on these self-described roles and views of participants in each harm case.

\subsection{Positionality Statement}
As researchers working in the sensitive space of understanding online harm and imagining interventions that may help address it, we briefly reflect on our position on this topic. 
All paper authors feel deeply concerned that online harm is a persistent social problem that disproportionately affects marginalized and vulnerable people \cite{duggan_online_2017}. Some authors also come from marginalized groups and are survivors of harm.
Our prior research on online harassment has helped us understand the inherent limitations of existing moderation mechanisms, which has shaped our desire to look for alternative approaches.
We celebrate the growing interest in applying alternative justice theories \cite{bell2008and,hurwitz2005explaining,kim2018carceral,meyer1998history,wenzel2008retributive}, e.g., racial justice \cite{schoenebeck2020drawing} and transformative justice \cite{daly2001transformative} to reduce harm to victims, offenders, communities, and societies. 
We are particularly enthused by the success of restorative justice in addressing offline harm and its potential to provide agency and care for vulnerable groups who are often ignored and harmed in a punitive justice model. 

While we embrace the values of restorative justice and see its potential, we do not seek to advocate for restorative justice in this work uncritically. We all live in cultures where punitive justice is the dominant approach to addressing harm, and we have actively worked to learn about restorative justice through research and by practicing its values in our own lives. This education and experience have shaped our interpretations of and understanding of the implications of our findings. At the same time, we acknowledge that restorative justice processes alone cannot address deep-rooted, systemic cultural and social issues such as racism and sexism (see transformative justice \cite{daly2001transformative}). 

We recognize that restorative justice is voluntary, and forcing this process may further harm the victims. Thus, we attempt to present an impartial account of our participants' perspectives on the victim-offender conference and restorative justice principles. 
Designing and implementing tools for restorative justice practices, and experimenting with these practices in online communities, are crucial areas for further understanding the utility of restorative justice in the online context.


\section{Findings: Current Moderation Models Through a Restorative Justice Lens} \label{current}
We use the three restorative justice principles (see Section 1) to understand the experiences of victims, offenders, and the community during instances of online harm within the current content moderation landscape. With our sampling methods, we included two victim-offender pairs in our study --- P17 and P25 as well as P22 and P24. To illustrate our findings, we present a relevant harm case from our data in each section. 

We warn readers that the following sections contain offensive ideas and language. However, we believe that such sensitive content can help readers understand the nature of online harm.

\subsection{Victims Have Limited Means to Heal from Past Harm or Stop the Continuation of Harm}
Restorative justice centers on victim needs. In offline restorative justice processes, victims usually share their stories, receive emotional support, and provide input on what is needed to repair the harm. The process aims to help them heal from harm and stop harm continuation~\cite{zehr2015little}. Our analysis indicates that the main tool available to online victims to address harm is reporting their offenders to moderators or the moderation system. 
However, it also shows that reporting such incidents does not effectively address their need for emotional healing or prevent future harm. 

To demonstrate the harm participants receive in online gaming communities, we first relate an example in Case 1.

\begin{longtable} [H] {|p{13.5cm}|} 
\hline

\textbf{Case 1}\\
    P13 teamed up with two previously unknown players to play an Overwatch game. She said \textit{hello} in the voice chat but instantly regretted it: \textit{``These guys started saying `Is that a girl damn are you cute,' and making jokes.''} P13 chose to remain silent, but she soon received a lurid threat: \textit{``They got harsh and said, `If you don’t respond I will take out my dick and slap you.'''}\\ 

    \hspace{5mm} P13 confronted those players, but the players ``\textit{all acted like it wasn't a big deal.}'' They continued to make jokes about her and complained about her gaming skills after they lost the first round. P13 decided to leave the game but continued monitoring the in-game chat: ``\textit{They were complaining about me, saying girls shouldn't play games. It makes me feel so nervous, I started to cry.}''\\
\hline
\end{longtable}

Several participants told us that they frequently experience harm, such as offensive name-calling or sexual harassment, on Overwatch or Discord. We find that such cases are often related to structural issues such as sexism, misogyny, transphobia, and homophobia---offenders often target their victims' gender or gender identity. Victims also received negative comments about their gaming skills. Women and gender non-conforming gamers sometimes experience compounding harms due to both of these patterns. P11 offered some examples of the comments she had heard while gaming: 

\begin{quote} \textit{
```Go make me a sandwich, calling me a bitch, telling me that I should go make food for my husband, I probably have kids and cats [\ldots] getting called bitch, slut, whore, cunt, just every derogatory name for women in the book.''
} \end{quote}

On both Overwatch and Discord, the main tool for victims to address harm is reporting to the moderation system or the moderators. However, victims often felt left out of the decision-making process after reporting a harm case and did not find the process helpful for healing from the harm. Several participants told us that after reporting, they either do not hear back or receive a vague message that indicates the moderation decision but offers no procedural details: \textit{``[The moderation system] just tells me that `something' happened, and my report led to that happening''} (P11). P9 believes that the moderation system is intentionally opaque to gamers: ``They don't want players understand[ing] the system.'' 

In addition, though reporting the harm may result in punishment of the offenders, it does not directly help victims heal from the emotional impact of the harm. P11 talked about the emotional toll of the incident: \textit{``I've cried about it a few times [\ldots and I told my friend], `I don't even feel like I want to live in a world where there are people like that.'''} 

Many victims indicated that though they tried to report their offenders, they continued to be harmed in the community, where a culture of harm is prevalent. Even if they do not encounter the same offender again, incidents of harm are so frequent that they come across new offenders repeatedly: \textit{``At least 5 out of 10 games, I get somebody saying some sort of crude comment about sexually harassing me''} (P25). When those experiences of harm accumulate, victims anticipate being frequently subjected to harm, and they accept that there are no effective ways to address it. P9 said, \textit{``[Harm] happens frequently enough that you just get used to it.''} P11 said, \textit{``I feel pretty helpless that I have to endure that every time I play a game.''} P12 noted that reporting itself becomes labor intensive: \textit{``[Harm] happens so often. I wouldn't want to report every single person I've talked to.''} 

Victims indicated that they need to consciously avoid harm in the community, which impacts their gaming experiences. For example, some participants stopped using voice chat out of fear that it would reveal their gender, even though communication with teammates is important for winning. P13 fears engaging with strangers after having experienced ongoing harmful behavior: \textit{``I will never play a game without a friend I trust, or talk in the game without my friends around because I don't trust that there will be even one person that will defend me when it gets bad.''} Some participants also leave a game or Discord community where they have experienced harm.

In sum, these findings show that the intensity and frequency of online harm can substantially impair users' online experiences and cause long-lasting emotional damage. Although the current moderation systems on platforms like Overwatch and Discord offer socio-technical mechanisms like reporting to address cases of online harm, the goals, and the lack of transparency and follow-up inherent in these mechanisms, do little to meet victims' need to heal. In addition, the current reporting mechanisms do not substantially reduce offenses within the gaming community because they do not effectively change the culture: even when participants can avoid the original offender, they continue to be harmed by new ones.

\subsection{Offenders Are Not Supported in Learning the Impact of Their Actions or Taking Accountability} \label{offender_section}

In a restorative justice view, offenders can address harm by acknowledging their wrongdoing, making amends, and changing their future behaviors. Through practices like victim-offender conferencing, offenders learn the impact of their actions by listening to the victim's side of the story. Afterward, they can repair the harm and learn through actions, such as acknowledging harm (e.g., apologizing), taking anger management courses, or doing community service \cite{zehr2015little, daly_restorative_2002}.

As our participants report, current moderation approaches often embody graduated sanctions \cite{kraut2012building}. That is, 
moderation teams in Overwatch and Discord tend not to ban offenders permanently after their first offense. Instead, they use more lenient punishment first to give offenders a chance to change their behaviors. Our interviews show that many Discord moderators have elements of a restorative mindset: they want to help offenders learn their wrongdoing and change their behavior by giving them second chances and providing explanations for their sanctions. Several Discord moderators explained their rationale for graduated sanctions as ``giving people a second chance.'' As moderator P14 noted, \textit{``We don't want [the moderation decision] to be a surprise [\ldots] and we will actually really want to encourage them to improve.''} Some Discord moderators also provide explanations of their moderation decisions. Such messages typically explain the rules and consequences of breaking them and can be posted by a moderator or a pre-configured bot.

Though some moderators have a restorative mindset, the punitive moderation system and rule-based explanations they rely on may fail to provide learning and support for offenders. They may not effectively stop the perpetuation of harm. We show an example through Case 2.

\begin{longtable} [H] {|p{13.5cm}|} 
\hline
  \textbf{Case 2} \\
    P14, a Discord moderator, described a case where moderators in her community gave an offender multiple chances to reform her behavior, but the offender was reluctant to change.\\

    \hspace{5mm} A transgender woman in P14's community had negative experiences with cisgender men in her life and ranted publicly in the community about her hatred of all men. Though the majority of the community are female and LGBTQ gamers, P14 strives to create an environment that is friendly to its cisgender male members: \textit{``[Cisgender men] are not allowed to say horrible things about the female and trans plus members, [so] in return we expect the same courtesy.’’}\\
    
    \hspace{5mm} The moderators warned the offending user several times. P14 reflected, \textit{``One of our mods would go in and be like, `Hey, just so you know, this is not okay here. I'm going to remove your message and just don't do it again.’ She would respond with, `Oh yeah, got to protect the cishet [ed. cisgender and heterosexual] men from the trans people.’’’}\\
    
    \hspace{5mm} This user was temporarily banned after several warnings. She then messaged moderators, telling them that they \textit{``don't know anything about oppression''} (P14). The moderators defended themselves, arguing that they understood oppression, and directed her to the community rules. Subsequently, the moderators stopped engaging with this member. \\

    \hspace{5mm} The temporary ban caused the transgender woman to lose her gender pronoun tag, which demonstrated her gender identity to the community. When she realized this had happened, she swore at the moderators, leading them to permanently ban her from the community.
    \\
    \hline
\end{longtable}

In case 2, the moderators made efforts to negotiate with the transgender woman, hoping that she would change by giving her multiple chances and providing explanations for their sanctions. However, their explanations mainly emphasize that her actions violated the rules and were prohibited, and the moderation decisions are all punitive. We argue that this model directs offenders' attention away from the victims and does not further their understanding of the impacts of their actions on the victim. Instead, the punishment, warnings, and restating of rules position the offenders against the moderation system and in defense of their own behavior. 

Restorative justice recognizes that often, offenders of harm may have been the victims of harm in other cases. As activist Mariame Kaba says: \textit{``No one enters violence the first time by committing it.''} \cite{happening?_2019}. In case 2, the harm that the transgender woman caused is a reaction to the harm she experienced elsewhere from cisgender men. Similarly, several offenders in our sample revealed how they had been constantly harmed in gaming communities: \textit{``I have been called countless names that are vulgar, offensive, and stuff like that'' (P21).} Restorative justice practices assume that although the past harms experienced by offenders do not absolve them of their responsibility for committing an offense, it can be hard to stop offensive behavior without addressing their sense of victimization through support and healing. Punishment, on the contrary, usually reinforces the sense of victimization \cite{zehr2015little}. 

The moderators in case 2 stopped the transgender woman from sharing her story. 
Instead, she received a denial, which worsened the damage and led her to defend herself more aggressively. In addition, when harm such as discrimination is a systemic issue in the gaming community or the broader society, stopping the perpetration of harm may require work to change the culture in addition to work to change individual offenders. Facilitator P6 explained how a restorative justice approach would strive to provide support and look at the root cause instead of punishing a particular offender:
\begin{quote} \textit{
``Rather than just zooming in on that one case and trying to blame this individual for that action, acknowledging that [they were] also harmed by the community and pushed to act in that way [\ldots] it's a much harder conversation to engage with, but then, that takes away from saying that person is wrong. It's more about [\ldots] how do we understand this collectively, and then, how do we address this collectively?
''} \end{quote}

We also find that when punishment is used as the only tool to stop offenses, it loses its function when offenders are not actually punished. Several moderators, victims, and offenders mentioned the convenience of using alternative accounts once one account has been moderated: \textit{``It is so easy to make new accounts in Discord that `reporting' people don't really work''} (P1). P17 used to conduct multiple offenses in games. He pointed out the fallacy of regulating with banning while there is no cost to creating and using another account in the game:

\begin{quote} \textit{
``Everyone right now is happy with that system because, if I was a normal player [\ldots] I heard that [the offenders] got banned, I would be like, `Wonderful, they got banned. I'm never going to see them again,' but in actuality, when I got banned I'm going to say, `I don't give a fuck, I'm just going to log into my second account.''} \end{quote}

Despite these shortcomings, we found evidence that the current punitive approach contributed to maintaining the health of community content and stopped the continuation of harm. Moderators we interviewed told us that some offenders would stop their misbehavior after receiving a warning or would not return to the same community after being banned. However, when punishment is the \textit{only} means of addressing harm, it cannot always be effective. 

\subsection{Gaming Communities Create Challenges for Victims and Offenders to Address Harm}
Harmful actions disrupt relationships within a community and potentially affect all its members. Thus, it is important for community members to acknowledge the harm and participate in redressing it collectively \cite{zehr2015little}. Restorative justice defines a \textit{micro-community} as the secondary victims who are affected by the harm and the people who can support the victims and offenders to address the harm (e.g., family or friends) \cite{mccold2000toward}. In online gaming communities, the micro-community includes not only victims, offenders, and moderators but also bystanders, friends of victims and offenders, and gamers, more generally. 

However, we found that the relevant stakeholders had no shared sense of community during many instances of online harm. As a result, the harm was often considered ``someone else's problem'' and remained unaddressed. Additionally, when micro-community members got involved, they often created secondary effects that further harmed victims and offenders.

\subsubsection{Victims, offenders, and moderators do not have a shared sense of community}
Restorative justice appeals to the mutual obligations and responsibilities of all community members to each other as necessary to address harm. However, in current moderation systems, we found that victims, offenders, and moderators lack a shared sense of community.

Many victims in our sample relate to and show care toward other gamers the offenders might harm. For example, P11 said, ``\textit{I usually am just sad not for myself really, but just that other people have to deal with those people.}'' Some victims even care about their offenders and what may have led them to perpetrate harm. However, there is a lack of \textit{shared} sense of connection from the offenders, which makes it hard for them to care about their victims and may lead them to fail to see the impact of their actions on others. The anonymous and ephemeral nature of online conversations deters offenders from relating or caring about the community or the victims. P17 used to harm other gamers but has since reformed himself. Reflecting on his previous mindset as an offender, he pointed out: \textit{``[The victims’] day is legitimately ruined because of what [offenders] said, but these people aren't going to think about what they said twice [\ldots] They're not going to reflect because it doesn't affect them.’’} 

Additionally, Overwatch randomly pairs up gamers when they do not join as a team. As a result, the offenders do not need to interact with their victims after a game has finished. This absolves offenders from feeling accountable for their actions and in fact creates an environment where repeating harm comes at no cost.  P21, who has attacked others in Overwatch games, noted: \textit{``In a game where you know somebody for 20 minutes and you see that they're bad so you flame [insult] them, and then you just move on. You never see them again.''} 

As mentioned in Section 5.2, many gamers we interviewed have multiple online identities in games and do not feel particularly attached to one. P17 used to attack others online using his anonymous accounts. He reflected, \textit{``No one can recognize what accounts we're on so we can do whatever we want.''} Given the limited information and social cues online, it is harder for people to find mutual relationships and connections. P21 used it to justify offenders' behavior: 
\begin{quote} \textit{
        ``In games you don't know these people. You don't know their personalities. You don't know their backstory [\ldots] All you know is that they're doing bad [at gaming], and so that's what you use against them. And, there's no way of fixing that. And if you're getting offended by some of these words, then just mute them.''
} \end{quote}

On many Discord servers, moderators regulate a confined, public space --- the general chat. Several moderators told us they do not intervene in harm cases that happen outside this general chat. Moderator P4 noted: \textit{``For the most part, we just tell people we can't moderate things that happen outside of our community, and at that point it's on them to block people.''} For harm cases in the general chat, moderators do not help offenders and victims mitigate problems; instead, they punish the offender or ask both parties to resolve problems themselves. Moderator P19 would move contentious conversations to a private space: \textit{``I'll delete all messages that they've put through to each other, put them both into the group chat, ban them from talking in general or whatever and keep them in this private chat and get it all solved in there.''}

In a restorative justice view, a sense of community is essential for offenders to care about the harm they cause to the victims and may even stop them from conducting harm in the first place. However, we found that the anonymous, ephemeral nature of online interactions makes offenders careless about breaking relationships with victims. Additionally, the unclear distinction between public and private online spaces creates challenges for moderators to support victims, and the ensuing lack of support may leave victims vulnerable.

\subsubsection{The community creates secondary harm against the victims}
In some offline practices of restorative justice conferencing \cite{zehr2015little}, community members who care about the harm that occurred can choose to join the conference. Community members can support victims by listening to their stories, acknowledging the harm that occurred, and providing emotional support \cite{zehr2015little}. Analyzing our interview data, we found that due to the absence of such a victim-centered structure online, community members often create secondary harm. Though it takes courage for victims to share their stories, finding an audience that supports and cares about them can be difficult. Instead, victims may be challenged, humiliated, or blamed, which creates secondary harm. 

P12 was sexually harassed by a male member in the gaming community when she was underage. She chose to reveal the member's behavior during a voice chat when other community members were present:  \textit{``I felt like more people should know about it since most of the people in [the Discord community] are underage.''} Though P12 expected to get support from her friends, she received multiple challenges from them after this revelation: \textit{``Some people were on my side [\ldots] but some other people were on his side and said I just wanted attention, and I should have known he was just joking.''} Later, the man denied all accusations and accused P12 instead. According to P12, \textit{``He was trying to make me look like a bad person.''} P12 felt so unsupported and hurt by the community's reaction that she eventually chose to leave it.

Community members often ignore the harms they witness, which can further fuel harm. P13 was a victim herself and has witnessed harm in the Overwatch community. She talked about how she was disappointed by the bystanders who did not speak up for the victims: 
\begin{quote}\textit{
``Most people just think that the game is over; there is no point in trying to help even your teammates, and think they will just end the game and try again another game. Since it is online, most people don't realize that it can actually hurt people what someone says.''
}\end{quote}

Further, community members can feed into harm by rewarding offenders with attention and positive reactions. P17 talked about why he used to harm intentionally to get attention from friends: 

\begin{quote}\textit{
    ``All my friends found it funny. The reason I said those things was not for [the victim \ldots] It's for them [my friends] to laugh. It's for the reaction. It's the adrenaline rush of being the center of attention, you know?'' 
}\end{quote}

These findings suggest that community members are not neutral bystanders when harm occurs. They may create secondary harm for victims by ignoring the harm, challenging their stories, or encouraging offenders' behavior. In contrast, restorative justice enables the community to build a culture of responsibility and care, where community members collectively prevent and address harm.

\subsubsection{The community has a punitive mindset toward offenders}
Restorative justice encourages the community to facilitate change in offenders' behaviors. Community members can share how the harm has impacted them, express willingness to support offenders through the change, and acknowledge any changes offenders have made. However, in punitive settings, community members who disagree with the offenders’ behavior want to punish them, as is the case in online moderation systems. Case 3 shows an example of such an occurrence and its effects in Overwatch.

\begin{longtable} [H] {|p{13.5cm}|} 
\hline
    \textbf{Case 3} \\
   P17 and his male friend were gaming as a team with a female player, P25, whom they started to attack with sexist comments after losing the first round. P25 decided to record this incident and posted it on Twitter after the attack continued for a while: \textit{``I knew that it was something that a lot of women and a lot of people deal with daily almost within the community. I wanted to be able to show just how bad it can be.''} Many people showed empathy and support for P25, and to date, the tweet has received more than 100 retweets and 600 likes. \\
   
    \hspace{5mm} After P25 reported P17 and his friend through Overwatch, their accounts were temporarily banned. However, for P17, the account banning was not the most severe punishment; rather, it was the subsequent harassment and damage to his reputation due to P25's public revelation. People located P17's multiple social media accounts. He reflected, \textit{``I would get random messages throughout the day saying, `You should kill yourself,’ and death threats like, `I'm going to come. I will.'''} He abandoned his previous online identity completely: \textit{``I deleted my Twitter, deleted Instagram, deleted Discord, changed all my Overwatch account names.''} P17 later apologized to P25 and has since changed his behavior. However, P17 believed the incident changed his career path as a 16-year-old: \textit{``To be honest, I think that was the main reason I didn't try to pursue to go pro in Overwatch harder [\ldots] because of how much I had to do to get back the reputation.''} \\
    
    \hspace{5mm} P17 suffered bans and community condemnation, which stopped his offensive online behavior. However, no one supported him in taking accountability and changing after learning the impact of his actions except for  one gamer friend who reached out to him. This friend suggested that P17 interact more with people offline because he seemed emotionally detached from his online victims. At that time, P17 used to play video games all day. P17 took that advice, and as time went on, he began realizing the impact of his actions on others: \textit{``[By] talking to real people and interacting with them face to face so you can see their emotions [\ldots] now I can imagine them [people I attacked in games] sitting at their computer and just like crying [\ldots] and that's why I don't say these things.''} \\
\hline
\end{longtable}

In Case 3, victim P25 had a positive experience when sharing her experiences of harm on Twitter. However, most other gamers decided to punish P17 instead of telling him how his actions caused harm. As we mentioned in Section \ref{offender_section}, the punishment only further harmed P17: \textit{``It's just anxiety. That's what's constantly going through your head.''} This punishment stopped P17 from offending again, but he did not learn to care about his victims until his friend reached out and supported him. 

Additionally, though P17 apologized to P25, stopped those behaviors, and learned the impact of his action afterward, he did not have a chance to demonstrate his change to the community. He left the community and abandoned his career goal to become a professional gamer. This is not what his victim, P25, had wished would happen. She wanted the offender to have a chance to learn and demonstrate his change: \textit{``Nobody's perfect, everybody makes mistakes. We're all human  [\ldots] I don't think that just because you did one bad thing a year ago means that you just don't have a chance anymore.''} The community response also runs counter to restorative justice views, which maintain that offenders will be welcomed back to the community after their reform \cite{zehr2015little}.

In general, we found that some community members have a sense of responsibility and care about the victims and the harm that occurs inside their  community. However, how they address harm is largely shaped by the punitive justice values prevalent in the gaming community and broader society. Their actions of punishing the offenders can create further harm in the community  and do not help offenders who may choose to reform themselves.
\section{Findings: Online Restorative Justice Possibilities and Challenges}
We now discuss how the various gaming community stakeholders responded to the idea of using one common restorative justice practice, namely, a victim-offender conference, to address harm in online gaming communities. Our goal was not to measure the participants' binary response regarding their interest in attending the victim-offender conference or use it as a direct measure of the potential of online restorative justice practices. Rather, we sought to identify the essential prerequisites and potential obstacles associated with designing and implementing new forms of online restorative justice by encouraging participants to reflect on their needs and concerns as related in the victim-offender conference.

\subsection{Victims' Needs and Concerns for a Restorative Justice Process} \label{victim_needs}
Many victims wanted to join a victim-offender conference if the offenders were willing to repair the harm. Concurrently, they had concerns and doubts about the process, especially about offenders' readiness to attend the conference. In our sample, no participants mentioned punishing offenders as a desired outcome of the victim-offender conference. We describe these needs and concerns below.

\subsubsection{Some victims want to understand and communicate with offenders}
Some victims wanted to understand why offenders harmed them and communicate to the offenders how they were hurt. Victims observed that the harm against them occurred unexpectedly, and they could not rationalize the offenders' behavior, which resulted in a need to understand it. P19 said, \textit{``I just want to know what goes through their head at that point in time.''} P11 expressed her confusion and frustration:

\begin{quote}\textit{
        ``I just don't understand the motive and don't understand the reasoning. Do they have girlfriends? [\ldots] Or their sisters? [\ldots] Because, I'm a sister, I'm a girlfriend, I have men in my life that I love and they would never do that to me. So why would you do it to someone else's loved one? I don't know.''
}\end{quote}

Several victims wanted to tell the offenders specifics of how they were hurt. P23 said, \textit{``First and foremost, I would express my feelings, how I felt and how that hurt me.''} Some victims hoped their sharing would help offenders realize the impact of their actions. P13 said, \textit{``Maybe hearing how hurt and scared I felt or feel, they would change their perspective.''} As someone who had once harmed others, P17 believed that learning how victims feel about their harm was essential to his change: \textit{``If I knew how people felt in games when I made fun of them, I would probably never make fun of them.''}


\subsubsection{Some victims want an apology, an acknowledgment of mistakes, and a promise of change from the offenders}
When we asked victims what they needed to repair the harm, several mentioned that they would like the offenders to realize their mistakes and issue an apology to start their emotional healing. As P13 said, \textit{``Just their realization of what was wrong with a small apology would make me happy.''} 

Other victims hoped that the offenders would change their behavior and stop offending. For some victims, causing offenders to reform their behavior was paramount:

\begin{quote}\textit{
    ``I wouldn't want him to do anything personally. I just want him to understand what he did wrong and try to fix it so it doesn’t happen again with me or another person.'' (P12).
}\end{quote}

\begin{quote}\textit{
``I don't really care to see their ranks [in games] drop or anything; I just want them to change.'' (P13)
}\end{quote}

\subsubsection{Some victims have concerns about whether offenders are willing to repair the harm}
During interviews, several victims were willing to join restorative justice meetings if the offenders were ready to repair the harm. However, they were concerned about whether the offenders they encountered would meet this condition. Several victims had already attempted to reach out to offenders to resolve the issue but were ignored or dismissed by them. These victims did not think the offenders would be open to participating in a restorative justice conference. P22 said, \textit{``I wouldn't be opposed to speaking to him [the offender] again, but I mean, from previous history, it's going to be difficult to speak to him or be able to trust him again because of his actions in the past.''} Some victims also questioned whether the offenders would genuinely want to repair the harm even if they consented to join the conference. P21 worried that the conference is a \textit{``get out of jail free card.''} He said, \textit{``People could be as offensive as they want, then turn around and be like, `Oh, I'm so sorry. I won't do it again.'''}

Most victims who shared these concerns did not want to meet with offenders if they could not ensure there was a genuine desire to repair the harm. One exception was P13, who had a strong preference for offenders to hear her voice even if they might be unaccommodating: \textit{``Even if they [offenders] are aggressive, as long as I had someone I trusted there, I would think it would have the best chance at an outcome [for the offender] to hear my voice.''} In restorative justice, victims and offenders meet only when both parties are willing to repair the harm. The facilitator also acts as the gatekeeper to ensure that further harm is unlikely before victims and offenders can meet \cite{bolitho2017science}.

\subsubsection{Some victims believe the harm they experienced is systemic and addressing it requires long-term efforts}
While several victims hoped that the victim-offender conference could reform offenders' behavior, others thought the problems they faced had systemic roots that could not be addressed in a single meeting. P16, a transgender woman who was harmed by someone she believed was transphobic, did not want to meet with the offender because she did not believe the meeting could solve the problem:
\begin{quote}\textit{
    ``I mean, because it's being trans, there's just so much systematic oppression to it [\ldots] It just takes time, and it takes education. It takes advocacy [\ldots] I believe he [the offender] is somebody who probably will change if he's given the right education, the right information, but it's going to take time and it's going to take people becoming more and more accepting of trans people. That's not just a fix that's going to be fixed easily through Discord.’’
}\end{quote}

P25, a female gamer who was verbally attacked while playing Overwatch (Case 3), 
believed the offender's aggressive behavior was likely shaped by more than just what happened in the video game. As a result, she thought the problem could not be solved within gaming communities alone:
\begin{quote}\textit{
    ``It's more probably deeper rooted than just the video game. It probably seeps into family issues to schooling issues to the environment they've grown up in, etc. You can't really help that without doing more and being there in person.''
}\end{quote}

\subsubsection{Some victims want to move past the harm}
Some victims noted that the harm had already occurred and wanted to move past it. Several were emotionally exhausted by the harm. P20 already felt disappointed by the offender's reaction when he tried to negotiate: \textit{``I don't think it's worth trying to regain her [the offender] as a friend [\ldots] because things like this can happen again due to the rash behavior.''} P16 described her feelings as \textit{``I'm just kind of over it.''}

Some victims thought the harm they experienced was trivial and that they were not severely impacted by it, or that they could reduce the emotional harm through their own efforts. As a result, they wanted to move on with their lives. P8 said, \textit{``If it's more personal, then they should maybe apologize. But [\ldots] in this case it's not particularly serious.''} P9 gave offenders the benefit of the doubt and felt that moving past the event was an easier option for him: 
\begin{quote}\textit{
    ``Maybe the people that were talking inappropriately were just having a bad day or something, or maybe they were genuinely not a good person and rude. But either way, it just seems easier for me to move past it by literally moving past it.'' 
}\end{quote}

In sum, most victims have needs other than punishing the offenders. However, they want offenders to engage with such needs only when they are sure that offenders genuinely wish to address the harm. In addition, many victims feel that restorative justice approaches might not be sufficient or ideal to address systemic online harm.


\subsection{Offenders' Interest in a Restorative Justice Process} \label{offender_needs}
During the interviews, we simulated pre-conference practices of restorative justice by asking offenders a series of questions to help them reflect on their experiences of harm and discuss their willingness to join a victim-offender conference. We now  describe offenders' thoughts on the process.

\subsubsection{Offenders may want to repair their relationships with the victims}
One offender (P22) wanted to join the victim-offender conference to repair the relationship with the victim (P24). He admitted that he was emotional when committing the harm: 

\begin{quote}\textit{
``I have since learned to control my temper as I have gotten older. However, it can still be a problem from time to time. I have an extremely competitive and stubborn personality, so when something doesn’t exactly go to plan, it can be difficult for myself to accept and get over it.''
}\end{quote}

P22 wanted to issue an apology to P24. They were organizing an Overwatch tournament together in a Discord community. P22 hoped that the meeting could help him maintain a professional relationship with the victim: \textit{``I would literally just settle for being civil with one another.''}

\subsubsection{Some offenders prefer the punitive process over restorative justice}
Several offenders (e.g., P17, P21) agreed that they should take full accountability for the harm they had caused. However, they preferred to receive punishment for their actions rather than join a victim-offender conference. In Case 3, P17 was banned for verbally attacking P25 and lost his reputation in the Overwatch community after P25 posted the video footage online. Though he acknowledged his mistake and apologized to P25 privately after the offense, he did not think he would have attended a victim-offender conference at that time. He described the process as \textit{``boring''}:
\begin{quote}\textit{
    ``I'm not going to go in with an open mind so nothing will get done anyways [\ldots] Would an immature 16 year old teenager who's rebellious against his mom, he doesn't want to do the freaking dishes, do you think he'll want to sit in a call or a meeting with the person that he just harassed for the last 20 minutes and figure it out? My answer is no.’’ (P17)
}\end{quote}

We found that these offenders' notion of justice aligns with the tenets of punitive justice. They believed they deserved punishment and did not trust the restorative justice process to help them achieve a better outcome. P17 said, \textit{``I definitely fucked up a little bit, so I deserve the punishment.''} P21 was banned for insulting others in a game. He was aware of his wrong-doing and expected to get punished even if he joined the meeting, so he wanted to go through the punishment directly: \textit{``If you know for a fact you're in the wrong, then there's no point in even talking, because you're going to get banned anyway.''} 

Both P17 and P21 were under-aged when they committed the harm. Facilitator P5, who works in a middle school, noted that the resistance P17 and P21 expressed is common in offline practices. P5 argued that the end goal of restorative justice is not to punish offenders but to help them take accountability for their actions and change future behavior. The pre-conference is a chance for facilitators to introduce restorative justice and talk about how it may benefit the offender. In school settings, P5 noted that the support for restorative justice from community members, including parents and teachers, encourages teenagers to be more open to the process: 
\begin{quote}\textit{
    \textit{``There's parents' support [\ldots and] there's so much research around how restorative justice works in a school and is incredibly beneficial for the students; it's an incredible healing process for the entire school community that a lot of [school staff] are willing to buy in.'' (P5) }
}\end{quote}

\subsubsection{Some offenders do not fully acknowledge their role in the harm}
Several offenders wanted to use victim meetings as a chance to justify their behavior. Those offenders believed that their victims behaved improperly or had hurt their interests in the first place. For example, P20 argued that the supposed victim lied about the situation, and he was wrongfully banned for defending himself. He hoped the victim could come to the meeting without preparation so he could challenge her by surprise: \textit{``They'll present the evidence right on the spot so that she doesn't have time to think or lie or really erase any evidence.''} This group of offenders agreed that they should take partial accountability for the harm they caused. However, they primarily wanted to use the meeting to hold the other party accountable and/or alleviate their own punishment:
\begin{quote}\textit{
``Why is it I'm being reported and banned after saying one thing, but yet you're not getting any punishments [for saying many]?''(P19)
}\end{quote}

As noted in Sec. 5.2, the current moderation systems lacks a means to hold offenders who use multiple anonymous accounts accountable for their actions. When offenders can conduct harm without receiving any consequence that they care about, participation in restorative justice becomes additional labor instead of an alternative to receiving punishment. As P21 said, \textit{``People aren't going to apologize. This isn't the real world [\ldots] If you get banned, they'll just go play another game.''}

In sum, we found one offender in our entire sample who wanted to repair his relationship with the victim, but the mindset of most offenders aligns with punitive justice. Offenders who acknowledge their wrongdoings think that they deserve punishment and do not believe that restorative justice can lead to a better outcome. Those who do not fully acknowledge their wrongdoings want to appropriate the victim-offender conference into a punitive process that holds victims accountable. These views reflect offenders' emphasis on the consequences of their actions on themselves rather than on  harm reparation.




\subsection{Moderators' Views on Implementing Restorative Justice Process in Their Communities} \label{moderation_needs}
We next discuss Discord moderators' attitudes toward implementing a restorative justice process in their communities. Most moderators agreed with the values of restorative justice, and some of their moderation practices overlapped with its  practices. At the same time, they expressed concerns about adapting the moderation process to the restorative justice model. Additionally, several moderators had already attempted restorative justice in their communities but received push-back and challenges from other moderators and community members. 

\subsubsection{There are elements of restorative justice in the current moderation practice on Discord, but for different purposes}
We find that both moderators and facilitators talk with victims and offenders when handling harm cases, but the issues addressed and the end goals differ. 
In offline restorative justice practices, facilitators talk with victims or offenders in pre-conferences, where they ask questions to determine what is needed to repair the harm \cite{pranis2015little}. Some Discord moderators also speak to victims and offenders before making a moderation decision, but their goal is to make informed decisions on how to punish offenders. 

A restorative justice pre-conference happens after fact-finding and focuses on emotions, impact, and the need to repair harm, and the facilitator shows support and empathy throughout the process \cite{van2016overview, bolitho2017science}. On the other hand, the conversation by moderators focuses on facts and evidence, with the moderators acting as judges. As moderator P3 described, \textit{``We will go and speak to whoever was reporting them [offenders] and speak to people who were involved, and try and get a feel for what actually happened and make a call from there.''}

In addition to the process of talking with victims and offenders, both restorative justice and the current moderation system aim to understand offender behaviors beyond the current harm case of interest. Restorative justice situates offenders in their life stories. As noted in Case 2, one life story would be that the transgender woman who offended cisgender men in the Discord community had had negative experiences with cisgender men in her life. A life story can help offenders find their triggers of harm, and the community could then provide support to help them heal \cite{karp_restorative_2001}. On the other hand, Discord moderators keep logs of past offenses for all users in the community. When an offender commits harm, the moderators review the logs as a reference to determine the proper punishment. As Discord moderator P2 explained, \textit{``We keep logs of all moderation actions that have been taken against any individuals. And so we always check those before handing out any issues [moderation decisions].''} 
Because of the graduated sanctions mechanism, the punishment is often heavier for offenders with past offenses. 


\subsubsection{Moderators' power may hinder restorative justice process} \label{Position_of_power}
Though we may think that the moderators' role is closest to that of facilitators, we find that the power moderators hold may impede restorative justice process. Moderator P7 has a work background in restorative justice, and he tried to implement pre-conferencing with offenders in his Discord community. He believed that he failed to reach desired outcomes with offenders because of his position of power. P7 banned an offender for cheating about his game rank to win. He conducted a pre-conference with him, where he wanted the offender to learn the impact of his actions on the victims he cheated on. However, the offender expressed the wish to get the ban revoked by offering professional Overwatch courses to P7 as an Overwatch coach. P7 was disturbed by the answer: \textit{``It was concerning, right? Because he's answering in a way to try and please me.''} P7 believed that the removal of power from online facilitators is essential for authentic sharing: 

\begin{quote}\textit{
``I don't want that kind of attitude of when people go into a performance they think of like, `Oh, I have to do well now, because they [moderators] have the decision-making power to remove me from [the Discord server].'''
}\end{quote}

Facilitators also pointed out that while moderators have the final right to interpret what has happened and who is right or wrong, restorative justice practices seek to give agency to victims and offenders. Moderators' power may create prejudices against victims or offenders and reduce their agency in the restorative justice process:

\begin{quote}\textit{
``As facilitators I'm never like, `Oh I take side with this story.' I'm always multi-partial, I hold the stories and then it's up for the individuals to figure out what's right to move forward.'' (P5)
}\end{quote}

Though removing punitive power from facilitators may be important, several moderators showed reluctance to let go of power. They worried that giving users agency may lead to unfair and biased outcomes because users may pursue an outcome that aligns with their own interests:

\begin{quote}\textit{
``They [the victims] are going to use it as kind of a tool to punish people that they don't like.'' (P25)
}\end{quote}

\begin{quote}\textit{
``[The community members in the conference] may initiate a witch hunt or just try and protect their friend group.'' (P3) 
}\end{quote}

These moderators' concerns are valid: it requires labor and skill from facilitators to address those potential issues. In offline practices, the facilitator is an essential role that maintains a power balance between victims and offenders, for instance, by ensuring they have equal opportunities to express their opinions. \cite{pranis2015little}.

\subsubsection{Some moderators think the labor of restorative justice is disproportional to its gains}
The pre-conference and conference processes require significant labor from facilitators \cite{pranis2015little}. Though a facilitator is usually paid in offline settings such as schools and prisons \cite{karp_restorative_2001, johnstone_handbook_2013}, Discord moderators are volunteers. Several moderators expressed concerns about the labor required to implement restorative justice. Moderator P7 said, \textit{``A lot of [moderators] are volunteers and so the easy option is to just mute people or temp ban people or permanently ban people.''} Being a facilitator also requires knowledge of the restorative justice practice. P3 thinks they would need to receive additional training to become a facilitator: \textit{``We're not qualified for this [\ldots] I don't know how we could provide support, or how to make sure that if we give the support, it's beneficial to them.''}

We found that many moderators are devoted to maintaining a healthy community environment, and some spend hours handling a single harm case. Moderator P14 gave an example: \textit{``To ban somebody, we actually can have about five or six hours worth of meetings [\ldots] to make sure that our punishment fits the crime essentially.''} They were concerned with the restorative justice process because they were unsure whether the extra labor would make any changes. Several moderators believed that users who re-offend multiple times have malicious intentions, so it is not worth spending more effort on them and helping them change. P1 said, \textit{``[Those harms] aimed at our identity (women, queer, trans, etc.) are often by people who enjoy calling us names. So, I don't really see the point of giving these people more opportunities to be prejudiced bigots.''} Similarly, P20 thinks offenders \textit{``can't be as civil''} in an online restorative justice conference compared to offline: \textit{``I do agree that having a talk and communicating would be nicer, but [\ldots] it generally doesn't go as well as in real life.''}

\subsubsection{Some moderators experienced challenges moving from individual restorative justice practices to institutional buy-in}
Several participants in our sample had an education or work background related to restorative justice and had attempted restorative justice practices in their gaming community. As noted in Sec. \ref{Position_of_power}, P7 conducted a pre-conference with an offender on the Discord server he moderates. P25, an Overwatch gamer, facilitated one victim-offender conference with two friends who fought during an Overwatch game. In the conference, the two friends acknowledged the impact and apologized to each other.

P7 and P25 independently initiated the restorative justice process. However, P7 believed in the importance of engaging the moderation team and other community members to practice restorative justice: \textit{ ``It takes more than one person to effectively execute restorative justice. For me, I have had a lot of roadblocks when it comes to trying to implement the system [alone].''} In offline scenarios, buy-in at the institutional level (e.g., schools, neighborhoods, workplaces) is important. Institutions can officially establish restorative justice as an alternative to the established punitive justice system \cite{johnstone_handbook_2013} and have resources to hire facilitators and remove or mitigate  punishment for offenders who successfully pursue the restorative justice process \cite{karp_restorative_2001}. Community members also gradually familiarize themselves with restorative justice and support victims and offenders in the process \cite{karp_restorative_2001, bazemore1999restorative}.

It is difficult for an online community to endorse restorative justice when the established culture and systems are punitive, and no examples demonstrate its effectiveness. As a moderator, P7 promoted restorative justice in his moderation team but failed: \textit{``You've got a traditional model and there's no real examples to demonstrate the capabilities of this [restorative justice].''} P18, a head moderator in P7's team, reflected on people's reactions to P7: \textit{``He (P7) kind of just mentions it [restorative justice], tries to explain it and then everyone gets confused and they kind of step back.''} Similarly, P8 tried to promote an alternative moderation system in Overwatch but could not get support from the gamers and Overwatch staff she talked to. She shared how people responded: \textit{``[People said that] the current system worked. It wasn't perfect, but it worked, and our system was new and untested.''} 
Despite such impressions, it is important to consider how we evaluate models for responding to interpersonal harm. For instance, have the victims' needs been met? Was harm repeated? Do offenders recognize the impact of their actions?  

In our interviews, we found it hard for people to imagine alternatives to the current moderation system. When we asked participants about their needs beyond having offenders banned or we introduced restorative justice to them, they often 
found it difficult to imagine the alternatives:

\begin{quote}\textit{
``It's not something I've thought about before.'' (P2)
}\end{quote}

\begin{quote}\textit{
``I'm not really sure what else you could do. Banned is the thing that everybody's always done.'' (P9)
}\end{quote}

In sum, many moderators spend time implementing practices that, on the surface, resemble restorative justice (e.g., talking with victims and offenders) but serve punitive purposes. In addition, the prevalence of intentional harm and the wide adoption of punitive models create challenges for communities that want to adopt a restorative justice approach.

\section{Discussion}
As noted throughout the paper, current content moderation systems predominantly address online harm using  a punitive approach. Analyzing through an alternate lens of restorative justice values, we found cases where this approach does not effectively meet the needs of victims or offenders of harm and can even further perpetuate the harm. Restorative justice provides a set of principles and practices that have the potential to address these issues. Table \ref{table:2} compares our sample's punitive content moderation approach with a restorative justice one. As the table shows, the latter provides alternative ways to achieve justice from the perspectives of victims, offenders, and community members, who are often absent in current content moderation.

\begin{table*}
\setcounter{table}{1}
\centering
\begin{threeparttable}
\caption{
Comparison of punitive content moderation with an approach based on restorative justice. 
}
 \label{tab:commands}
\def\arraystretch{1.5}
\begin{tabular}{ p{2cm}| p{5cm} | p{5cm}}

\midrule
 & \textbf{Punitive content moderation approach} & \textbf{Restorative justice approach} \\
\midrule
Victims & Victims are left out of the moderation process.	&
Addresses the victim's needs for reparation of harm, such as support and healing.	\\ \hline
Offenders & Offenders receive rule-centered moderation explanations and punishment.	&	
Encourages offenders to learn the impact of their actions and take responsibility to make reparations. \\ \hline
Community members & Community members may further harm or punish victims and offenders. & Engages community members in supporting victims and offenders and heal collectively.
	\\ 
\midrule
 \end{tabular}
  \label{table:2}
  \end{threeparttable}
\end{table*}

Applying restorative justice practices to the governance of online social spaces is not straightforward. Victims, offenders, and moderators currently have unmet needs and concerns about the process. 
In this section, we first discuss the challenges of implementing restorative justice. We then offer possible ways for interested online communities to implement the approach. Finally, we reflect on the relationship between punitive content moderation approaches and restorative justice.

\subsection{Challenges in Implementing Restorative Justice Online}
We now explore the potential structural, cultural, and resource-related challenges of implementing restorative justice online. We also discuss some possible ways to address them. 
 \subsubsection{Labor of restorative justice}
Restorative justice practices and the process of shifting to them will likely require significant labor from online communities. Restorative justice conferencing can be time intensive for all stakeholders involved. Before victims and offenders can meet, facilitators must negotiate with each party in pre-conferences, sometimes through multiple rounds of discussion. The collective meeting involves a sequence of procedures, including sharing, negotiating, and reaching a consensus. 

Stakeholders, in particular facilitators, must expend emotional as well as cognitive labor. In offline practices, facilitators are usually paid by host organizations such as schools \cite{gavrielides2017restorative}. However, the voluntary nature of moderation on social media sites like Discord means that online facilitators may be asked to do additional unpaid work. This issue can be particularly salient when moderators already expend extensive labor with a growing community \cite{roberts_behind_2019, dosono2019moderation, steiger2021psy}. Labor is also involved in training the facilitators. Unlike punitive justice, restorative justice is not currently a societal norm that people can experience and learn about on a daily basis. Aspiring facilitators need additional training to learn restorative justice principles and practices and implement them successfully to avoid creating additional harm. 

We estimate that resources for addressing the aforementioned labor needs could be attained in both a top-down and a bottom-up fashion. A top-down process could require resources from companies that host online communities. 
There is precedent for platforms making such investments; in recent years, social media companies such as Discord have hosted training for its community moderators \footnote{Discord Moderator Academy. https://discord.com/moderation}. A bottom-up process could engage users with preexisting knowledge about restorative justice to first introduce the process in their communities and gradually expand the restorative culture and practice from there. 
In our sample, two moderators attempted or practiced online restorative justice within their own communities; they showed enthusiasm for expanding its  practice to other online gaming communities. It is possible that resources from companies \textit{and }practitioners could collectively begin to address the labor problem. 

Additionally, implementing online restorative justice requires\textit{ reallocationing} labor instead of merely adding labor. We found that many moderators we interviewed already practiced different elements of restorative justice. Some aim to support victims and give offenders a second chance but do not have the proper tools or procedures to achieve that. Other moderators have practices that embed elements of restorative justice, such as talking with offenders and victims. Rather than necessarily requiring new procedures, restorative justice requires a shift of purpose in existing processes -- from determining the point of offense to caring for victims and offenders. 

Importantly, if online restorative justice could stop the perpetuation of harm more frequently than punitive justice, it could \textit{reduce }the need for moderation labor in the long term. While research has shown that offline restorative justice has successfully reduced re-offense rates \cite{latimer_effectiveness_2005, morris1993giving, daly_restorative_2002}, evaluating the effectiveness of restorative justice practices in online communities is an important area for future work.



\subsubsection{Individuals' understanding of justice aligns with the existing punitive justice model}

Although people have needs that the current system does not address, we found that their mindsets and understanding of potential options often align with what the current system can already achieve through its punitive approach. As our research shows, many moderators and victims think that punishing offenders is the most or best they can do, and some offenders also expect to receive punishment. Community members also further perpetuate the punishment paradigm. This mindset is not only a result of the gaming community’s culture; it is pervasive throughout society, including in prisons, workplaces, and schools \cite{foucault_discipline_2012, king2009conservative}. 

Given the lack of sufficient information about and experiences with restorative justice, people may misunderstand and mistrust this new justice model. Some offenders in our interviews  
still expected to receive bans after the process, but restorative justice usually serves as an alternative to punishment, not an addition to it. Some participants wanted to implement alternative justice models in their own communities but received resistance from users who argued that the current moderation system works for them while disregarding its limitations for others. We found that such perspectives usually lead to quick rejections of the notion of implementing restorative justice online. 

Before people can imagine and implement alternative justice frameworks to address harm, they must be aware of them. Crucial steps in this direction 
are information and education. 
Helping people understand the diversity of restorative justice processes and how their aim is restoration instead of punishment may address their doubts and open more opportunities for change. This is especially important since an incomplete understanding of restorative justice may cause harm in its implementation. For example, enforcing ``successful outcomes'' may disempower victims and result in disingenuous responses from offenders. 
Adapting restorative justice to online communities may require changes in the format and procedure of how harm is handled, but prioritizing its core values should help avoid additional unintentional harm.


Restorative justice has developed and rapidly evolved in worldwide practice \cite{zehr2015little, daly2001conferencing}. Future research can build on and expand restorative justice beyond the three principles and the victim-offender conference. In addition, people need to experience it to understand it, adapt it to their needs, and learn about its effectiveness. By experiencing it offline, several participants in our sample came to see it as a natural tool adaptable to online communities. 

In future work, we plan to collaborate with online communities to implement and test restorative justice practices. We want to pilot online restorative justice sessions and run survey studies to understand the types of harm and types of processes most likely to benefit from these practices. Building on that research, we aim to provide more precise empirical guidelines about how restorative justice can be embedded in moderation systems based on the socio-technical affordances of various online communities. To manifest a future of restorative justice, \textit{``We will figure it out by working to get there.''} \cite{kaba_2021}.

While this research focuses on the restorative justice approach, it is not the only alternative and has its limitations. As some of our participants mentioned, many harms are rooted in structural issues such as sexism and racism. \textit{Transformative justice} \cite{daly2001transformative} and \textit{community accountability} \cite{molyneux2012community} are frameworks of justice developed to address such issues. \textit{Procedural justice} emphasizes the importance of a fair moderation decision-making process \cite{hou2017factors} and is a key objective in restorative justice practices \cite{barnes2015restorative}. Future work should explore the potential of these different justice frameworks to address online harm.

\subsubsection{Offender accountability}
Another challenge may be in motivating offenders to take accountability for their wrongdoing -- a persistent moderation problem regardless of the justice model implemented. In our interviews, we found that given the finite scope of moderation in many contexts and the limits in technical affordances of online communities, offenders can often easily avoid punishment. The harm may happen in a place without moderation or clear rules of moderation, e.g., when harm occurs during a private Discord chat or across multiple platforms. Some participants also noted that having multiple identities/accounts in Overwatch or Discord is easy. Thus, when punishment is ineffective, punitive justice may also lose its effectiveness.

Punishment is not the only form of holding offenders accountable. Restorative justice believes that people are also connected through relationships. Our interview data, as well as restorative justice literature, suggest the importance of a sense of community. If offenders perceive themselves as members of a shared community with victims, they will be more likely to take accountability for addressing harm \cite{pranis2015little, karp_restorative_2001}. 
However, in our interviews, we find that there exists a lack of sense of community. Offenders may not expect to meet victims again or hide behind multiple anonymous accounts. Moderators typically moderate a confined space of general chat, which can leave harm unaddressed in any place outside.

Therefore, it is vital that we inquire into what accountability means to the community and how to hold people accountable within the current moderation system. If offenders can simply avoid any punitive consequences of conducting harm and do not feel a sense of belonging to the community where harm occurs, it would be challenging to engage them in any process --punitive or restorative-- that holds them accountable. 


\subsubsection{Emotion sharing and communication in restorative justice}
The limited modes of interaction and the often anonymous member participation in online platforms may influence the effectiveness of restorative justice. Many online interactions are restricted to text or voice, prohibiting victims and offenders from sharing emotions, and may give rise to disingenuous participation. Emotional engagement by victims and offenders is essential for empathy and remorse \cite{strang2003repairing}. Face-to-face sharing lets victims and offenders see each other's body language and facial expressions. 

Implementing offline restorative justice includes a series of physical structures and meeting procedures to elicit genuine, equal sharing. For example, participants sit in a circle; individuals who speak hold a talking stick; there are rituals at the beginning of the conference to build connections among participants and mark the circle as a unique space for change \cite{pranis2015little}. Those rituals for emotion sharing are hard to replicate in the online space. For example, if an offender messages an apology through text, it can be harder to discern a genuine apology from a disingenuous one. 



The issue of computer-mediated communication and emotion-sharing has been long discussed in the HCI and CSCW literature. In recent years, increasingly more advanced technologies have been developed to facilitate civic engagement and communication in online systems. For example, Kriplean et al. built a platform, ConsiderIt, to support reflective interpersonal deliberation \cite{kriplean2012}.
REASON (Rapid Evidence Aggregation Supporting Optimal Negotiation) is a Java applet developed to improve information pooling and arrive at consensus decisions \cite{introne2008adaptive}. 
Many researchers have attempted to model human-like characteristics and emotional awareness in chatbots \cite{adikari2019cognitive,spring2019empathic}. 

In the context of the restorative justice approach, Hughes has developed a tool, Keeper, for implementing online restorative justice circles \cite{hughes2020keeper}. Such existing systems can be leveraged to develop advanced tools that facilitate emotion-sharing and communication in online restorative justice processes. Online platforms such as Overwatch and Discord could add such tools to improve emotional sharing and communication, necessary conditions for implementing restorative justice.

\subsection{Applying Restorative Justice in Online Moderation} \label{RJ_possibilities}
We now discuss possible ways to adapt  current moderation practices to implement restorative justice online. While we have used victim-offender conferencing as a vehicle to interrogate the opportunities and challenges of implementation, restorative justice includes a set of values and practices that extend beyond the conference. 
It is important to meet online communities and platforms where they are and design restorative justice mechanisms based on their resources, culture, and socio-technical affordances. For example, compared to Overwatch, Discord provides more flexibility in communication and ways of maintaining social connection after a game. Some Discord servers have a greater sense of community and moderation resources than others. We suggest that online communities or platforms can begin with partial restorative justice practices that involve only a few stakeholders (e.g., pre-conferencing) or adapt some moderation practices to embed restorative justice values (e.g., moderation explanations) and implement victim-offender conferencing when its preconditions are met. 

\subsubsection{Embed restorative justice language in moderation explanations}
Moderators can embed restorative justice language in explanations of moderation decisions. Prior work has shown that providing such explanations can reduce re-offense in the same community. However, many moderated offenders dismiss such explanations and continue re-offending \cite{jhaver2019survey, jhaver2019transp}. Our research shows that explanations often focus on describing community guidelines or presenting community norms to the users, not encouraging offenders to reflect on the impact of their actions. These explanations usually indicate the actual or potential punitive consequence, which may direct offenders' attention to the moderation system instead of their victims. 

We suggest a shift in language from a rule-based explanation to an explanation that highlights the possible impact offenders may cause to victims and supports the offender in taking accountability. Facilitator P5 provided an example of how she would communicate with the offenders if she were an online facilitator: \textit{``This post had this emotional impact [\ldots] this is how you've caused harm. This is the feedback from the community, and we want to work with you to create change.''}

Victims are often left out of the moderation process in the online communities we studied. However, some victims want information on the moderation process being used and results of moderator interventions.  While prior research has discussed how moderation transparency prevents offenders from re-offending \cite{jhaver2019transp}, our work highlights  the importance of providing information for victims in the moderation process since they are the ones with needs for support and healing. We suggest that a note of care may help victims feel heard and validated and help them recover. 

\subsubsection{Restorative justice conferences with victims or offenders}
In our interviews, we found that some victims or offenders may not be available or willing to meet and have a conversation to address the harm. When a collective conference is not possible, it is possible to apply a \textit{partial restorative justice process} that includes offenders or victims alone \cite{kuo2010empirical, mccold2000toward}. Some Discord moderators already talk with victims and offenders before making moderation decisions, a practice similar to restorative justice pre-conferencing. In offline pre-conferencing, facilitators ask questions to help victims and offenders reflect on the harm, providing them with opportunities for healing and learning. 

We propose that pre-conferencing provides opportunities to meet some of the needs our participants identified. For example, when an offender wants to apologize to the victim, the victim may not want to meet the offender but communicate their feelings through the facilitator. 

In offline restorative justice, pre-conferencing is a preparation step for a potential victim-offender conference. Thus, if both victim and offender are willing to meet with each other in the pre-conference, the moderators can organize a victim-offender conference, where both parties share their perspectives and collectively decide how to address the harm. Moderators have the responsibility to maintain a safe space for sharing, for example, to halt the process when harm seems likely to happen, ensure a power balance between the victim and the offender, and work with participants to establish agreements of behavior and values to adhere to throughout the process \cite{bolitho2017science}. In cases where restorative justice does not succeed, preventing the continuation of harm must be prioritized. Because facilitating these processes is difficult, we discuss the challenges and opportunities for moderators to facilitate in the following section.

A victim-support group is another form of restorative justice conferencing \cite{dignan2004understanding}. In our sample, many victims indicated a need to receive emotional support. We propose that online communities offer a space for victims to share the harm they experienced and receive support. Systems with similar goals have previously been built in online spaces. For example, Hollaback is a platform that allows victims of street harassment to empower each other through storytelling \cite{dimond2013hollaback}. Heartmob enables victims to describe their experiences of harm and solicit advice and emotional support from volunteer Heartmobers \cite{blackwell2017classification}. These platforms offer templates for how victim support systems can be built. However, support in these platforms is distant from where harm occurs, and it is also important to think about how online communities can support victims by motivating community members to provide support.


\subsection{Situating Restorative Justice in the Moderation Landscape}
We have illustrated possible ways to implement restorative justice in the current moderation system. Yet, we do not seek to advocate for restorative justice as a wholesale replacement of the current moderation approach. We propose that \textit{restorative justice goals and practices be embedded as part of an overall governance structure in online spaces}.  

Restorative justice conferencing should be used in select cases only because it is effective only when it is voluntary and the parties involved are committed to engaging. Our findings show that individuals have different conceptualizations of justice. Schoenebeck et al. also found that people's preferences for punitive or restorative outcomes vary with their identity or social media behaviors \cite{schoenebeck2020drawing}. It is thus important to attend to victims' and offenders' preferences in individual harm cases. 

A larger governance structure should also take into account what to do if restorative justice processes fail. For instance, if it is determined at the pre-conference stage that a restorative justice approach cannot be applied, actions such as removing access by muting or banning might be used to prevent the continuation of harm. This is consistent with offline restorative justice practices, where the community or court clearly defines the types of cases that go through a restorative justice process and the action to take if a restorative justice process is not possible \cite{gavrielides2017restorative, van2016overview}.


We estimate that whether an online community decides to apply restorative justice is a value-related question. While restorative justice and punitive justice processes share a  goal of addressing harm and preventing its perpetuation, they have different processes and orientations toward when justice is achieved. Content moderation---closer to a punitive justice approach---addresses harm by limiting the types of content that remain on the site and punishing offenders in proportion to their offense. In contrast, restorative justice  aims to assure that victims' needs are met, and offenders learn, repair harm, and stop offending in the future. Thus, the primary reason for applying restorative justice in moderation is not to achieve the current goals of effectively removing inappropriate content and punishing offenders but to benefit online communities using restorative justice values and goals. 

Seering et al. found that community moderators have diverse values regarding what moderation should achieve \cite{seering2020metaphors}. While some have a more punitive mindset and hope to be a ``governor,'' others align more with restorative justice values and hope to be ``facilitator'' or ``gardener.'' Thus, online communities must  reflect on their values and goals and decide on what mechanisms (e.g., punitive or restorative) help realize those values. Recent research has argued that social media platforms are responsible for incorporating ethical values in moderation instead of merely optimizing to achieve commercial goals \cite{helberger2018governing}. In particular, some researchers have proposed values and goals that align with restorative justice, such as centering victims' needs in addressing harm \cite{schoenebeck2020drawing, blackwell2017classification}, democracy \cite{seering2020reconsidering}, and education \cite{myers2018censored}. Our work adds to this line of research and envisions how restorative justice may benefit online communities in addressing some severe forms of online harm, such as harassment.

Finally, communities should be cautious about expecting or enforcing a positive outcome. Enforcing forgiveness from victims or expecting a change in offenders' behavior may undermine victims' needs and put them in a vulnerable place for forgiveness or induce a disingenuous response from offenders \cite{bazemore2015restorative}. Online communities should allow for partial success or no success without enforcing the ideal outcome, especially at the early stage of implementation when there are insufficient resources or commitments. Instead, they  may focus on how victims, offenders, and the entire community could holistically benefit from the process.



\section{Conclusion} 
In this research, we interviewed victims, offenders, and moderators in the Overwatch gaming community. We presented case studies that identified opportunities and challenges for using restorative justice in addressing online harm and discussed possible ways to embed a restorative justice approach in the current content moderation landscape. Much remains to be done to explore what restorative justice may look like in an online community context and when and how to implement this approach. We hope this work offers a valuable guide for designers, volunteers, activists, and other scholars to experiment with restorative justice 
 and related approaches in online communities.

\begin{acks}
We would like to thank Julie Shackford-Bradley and Joseph Seering for their valuable feedback. This research was supported by the National Science Foundation award IIS-1948067.
\end{acks}

\bibliographystyle{ACM-Reference-Format}
\bibliography{references, shagun-references, sijia-reference, niloufar-references}

\end{document}